\documentstyle[prb,multicol,float,aps,epsf]{revtex}
\newcommand{\blong}{\ifpreprintsty
                   \else
                   \end{multicols}\vspace*{-3.5ex}{\tiny
                   \noindent\begin{tabular}[t]{c|}
                   \parbox{0.49\hsize}{~} \\ \hline \end{tabular}}
                   \fi}
\newcommand{\elong}{\ifpreprintsty
                   \else
                   {\tiny\hspace*{\fill}\begin{tabular}[t]{|c}\hline
                    \parbox{0.49\hsize}{~} \\
                    \end{tabular}}\vspace*{-2.5ex}\begin{multicols}{2}
                    \fi}
\def\be{\begin{equation}}
\def\ee{\end{equation}}
\def\bea{\begin{eqnarray}}
\def\eea{\end{eqnarray}}
\begin{document}

\title{Determinations of upper critical field in the $c$-$a$ plane} 

\author{L. Wang, H. S. Lim, and C. K. Ong}

\address{Center for Superconducting and Magnetic Materials and 
Department of Physics, Blk. S12, \\
Faculty of Science, \\
National University of Singapore, \\
2 Science Drive 3, \\
Singapore 117542. \\
} 
\maketitle
\begin{abstract}
Within a continuous Ginzburg-Landau model for layered superconductors, two procedures 
are proposed to determine the upper critical field parallel to the $c$-$a$ crystal plane 
with an angle $\theta$ tilted from the $c$-axis. For an intrinsically layered superconductor, 
the upper critical fields for $\theta \rightarrow 90^\circ$, as determined by  
the two procedures, are consistent with each other and in reasonably good agreement with 
those determined by similar procedures suited for the parallel upper critical field 
($\theta = 90^\circ$). The profile of the order parameter obtained at $B_{c2}$ is 
Gaussian-like, indicating the plausibility of the procedures proposed.  
\end{abstract}
\begin{multicols}{2}
\tighten

\section{Introduction}

Theories within the Ginzburg-Landau (GL) framework are important for studies of 
properties of layered superconductors. Considering the layered structure of 
high T$_{c}$ superconductors (HTSs), Koyama {\it et al.}\cite{Koyama92} first proposed 
a continuous Ginzburg-Landau (CGL) model, in which the GL coefficients and 
the superpair masses are spatially dependent. Recently, we have proposed a similar 
CGL model and have applied it to a layered superconductor without 
a magnetic field\cite{Wang01a}, and to layered systems with a magnetic field 
parallel to the $a$- ($b$-) crystal axis\cite{Wang01b}. Details of the determinations of the 
parallel upper critical field were given in Ref.~\onlinecite{condmat1}. In this work, 
our CGL formulation is applied to layered superconductors immersed in 
a magnetic field parallel to the $c$-$a$ crystal plane. Two procedures shall be 
proposed to determine the upper critical field, which is a subject of long-term interest\cite{Maska01}. 

\section{Model}

We consider a layered superconductor comprising of alternating superconducting (S) 
and insulating (I) layers along the $c$-axis. 
The CGL free energy for the system is\cite{Wang01a},  
\blong
\bea
F & = & \int{d\vec{r}}\int{dz} \left[ \alpha(T,z) |\Psi(\vec{r},z)|^{2} +
\frac{1}{2} \beta |\Psi(\vec{r},z)|^{4} +
\frac{\hbar^{2}}{2M(z)}\left|\left( \frac{\partial}{\partial z}-
\frac{2ie}{\hbar}A_{z} (\vec{r},z)\right) \Psi
(\vec{r},z)\right|^{2} \right. \nonumber \\
& & \mbox{} + \left. \frac{\hbar^{2}}{2m(z)}\left| \left( \nabla^{(2)} -  
\frac{2ie}{\hbar}\vec{A}^{(2)}(\vec{r},z)
\right)\Psi(\vec{r},z)\right|^{2}
+ \frac{1}{2\mu_{0}}B^{2}(\vec{r},z) \right], 
\label{eq:freeE}
\eea
\elong\noindent
where $\vec{r}=(x, y)$ is the planar vector and $\vec{A}(\vec{r},z) = (\vec{A}^{(2)}(\vec{r},z), A_{z}(\vec{r},z))$ is the vector potential. $\Psi(\vec{r},z)$ 
is the superconducting order parameter and $B(\vec{r},z)$ is the internal magnetic field. 
$\beta$ is assumed constant\cite{Wang01a,Kleiner97}. The CGL condensation coefficient $\alpha(T,z)$ and the perpendicular and parallel effective masses, $M(z)$ and $m(z)$, 
are assumed as before\cite{Wang01a,Wang01b}: 
\begin{mathletters}\label{eq:coeficient}
\begin{eqnarray}
\alpha (T,z)& =&  \left[ \alpha_0 + \alpha_{1}\cos (2\pi z/D)\right] (1-T/T_{c}),  \label{eq:coeficienta} \\ 
\frac{1}{M(z)} & = & G_{0} + G_{1} \cos (2\pi z/D), 
\label{eq:coeficientb} \\
\frac{1}{m(z)} & = &  g_{0} + g_{1} \cos (2\pi z/D), 
\label{eq:coeficientc}
\end{eqnarray}  
\end{mathletters}
where $\alpha_0$, $\alpha_1$, $G_0$, $G_1$, $g_0$ and $g_1$ are the model parameters. 
$D$ is the size of the unit cell equal to $d_I/2+d_S+d_I/2=d_I+d_S$, 
with $d_I$ and $d_S$ denoting the thickness of the I and S layers, respectively. 
In the present simulation for layered superconductors, the intrinsically layered cuprate Bi$_2$Sr$_2$CaCu$_2$O$_8$ (Bi2212) is chosen as the modeling prototype as 
this superconductor has an explicitly layered structure due to its high 
anisotropy\cite{Farrell89}. Details of the calculation inputs for Bi2212 can be 
found in Ref.~\onlinecite{Wang01b}.

Let us now consider the case where an external magnetic field $B$ is applied 
in the direction tilted from the $z$-axis ($c$-axis) by an angle $\theta$ in 
the $z$-$x$ ($c$-$a$) plane. Taking the vector potential 
as $\vec{A}=(0, B(x\cos\theta-z\sin\theta), 0)$, the linearized CGL equation from 
the CGL free energy is obtained as follows, 
\blong\bea
-\frac{\hbar^{2}}{2M(z)}\frac{\partial^{2}}{\partial z^{2}}\Psi(x,y,z) -
\frac{\hbar^{2}}{2}\left[ \frac{\partial}{\partial z} \frac{1}{M(z)}
\right]\frac{\partial}{\partial z} \Psi(x,y,z)  \nonumber \\
\mbox{} - \frac{\hbar^{2}}{2m(z)}\left[ \frac{\partial^{2}}{\partial x^{2}}+\left(\frac{\partial}{\partial {y}}-2ieB(x\cos\theta-z\sin\theta)\right)^2 
\right] \Psi(x,y,z) + \alpha(T,z)\Psi(x,y,z)  =  0.
\label{eq:GL}
\eea
\elong\noindent
In a new coordinate system, 
\begin{eqnarray}
\left(
{
\matrix{
x^{\prime} \cr
y^{\prime} \cr
z^{\prime} \cr
}
}
\right)
=
\left(
{
\matrix{
\cos\theta & 0  & -\sin\theta    \cr
0          & 1  & 0              \cr
\sin\theta & 0  & \cos\theta     \cr
}
}
\right)
\left(
{
\matrix{
x \cr
y \cr
z \cr
}
}
\right),
\label{eq:xyz}
\end{eqnarray}
Eq.~\ref{eq:GL} becomes 
\blong\bea
-\frac{\hbar^{2}}{2M(x^{\prime},z^{\prime})}\left(-\sin\theta\frac{\partial}{\partial x^{\prime}}+\cos\theta\frac{\partial}{\partial z^{\prime}}\right)^2\Psi(x^{\prime},y^{\prime},z^{\prime}) \nonumber \\
\mbox{} -\frac{\hbar^{2}}{2}\left[ \left(-\sin\theta\frac{\partial}{\partial x^{\prime}}+\cos\theta\frac{\partial}{\partial z^{\prime}}\right) \frac{1}{M(x^{\prime},z^{\prime})}\right] 
\left(-\sin\theta\frac{\partial}{\partial x^{\prime}}+
\cos\theta\frac{\partial}{\partial z^{\prime}}\right) 
\Psi(x^{\prime},y^{\prime},z^{\prime})  \nonumber \\
\mbox{} - \frac{\hbar^{2}}{2m(x^{\prime},z^{\prime})}\left[ \left(
\cos\theta\frac{\partial}{\partial x^{\prime}}+\sin\theta
\frac{\partial}{\partial z^{\prime}}\right)^2
+\left(\frac{\partial}{\partial {y^{\prime}}}-2ieBx^{\prime}\right)^2 \right] \Psi(x^{\prime},y^{\prime},z^{\prime}) \nonumber \\
\mbox{}+\alpha(T,x^{\prime},z^{\prime})\Psi(x^{\prime},y^{\prime},z^{\prime})  =  0.
\label{eq:neweq}
\eea\elong\noindent
Assuming $\Psi(x^{\prime},y^{\prime},z^{\prime})= e^{iky^{\prime}}\Phi(x^{\prime},z^{\prime})$, it follows from Eq.~\ref{eq:neweq} that
\blong\bea
-\frac{\hbar^{2}}{2M(x^{\prime},z^{\prime})}\left(-\sin\theta\frac{\partial}
{\partial x^{\prime}}+\cos\theta\frac{\partial}{\partial z^{\prime}}\right)^2\Phi(x^{\prime},z^{\prime}) \nonumber \\
\mbox{} -\frac{\hbar^{2}}{2}\left[ \left(-\sin\theta\frac{\partial}{\partial x^{\prime}}+\cos\theta\frac{\partial}{\partial z^{\prime}}\right) \frac{1}{M(x^{\prime},z^{\prime})}\right] 
\left(-\sin\theta\frac{\partial}{\partial x^{\prime}}+\cos\theta\frac{\partial}
{\partial z^{\prime}}\right) \Phi(x^{\prime},z^{\prime})  \nonumber \\
\mbox{} - \frac{\hbar^2}{2m(x^{\prime},z^{\prime})}\left[ \left(\cos\theta\frac
{\partial}{\partial x^{\prime}}+\sin\theta
\frac{\partial}{\partial z^{\prime}}\right)^2
-4e^2B^2(x^{\prime}-x^{\prime}_0)^2 \right] \Phi(x^{\prime},z^{\prime}) \nonumber \\
\mbox{}+\alpha(T,x^{\prime},z^{\prime})\Phi(x^{\prime},z^{\prime})  =  0,
\label{eq:eq2DFinal}
\eea\elong\noindent
with $x^\prime_0=\hbar k/(2eB)$. For $0 \leq \theta < 90^\circ$, one may choose 
$x^\prime_0=0$\cite{Takezawa93}. 
 
For a given temperature $T$, the maximum magnetic field $B$ which satisfies 
Eq.~\ref{eq:eq2DFinal} gives a point on the $B_{c2}$-$T$ plot. Eq.~\ref{eq:eq2DFinal} 
shall be numerically solved subject to the following boundary conditions\cite{Takezawa93},
\begin{mathletters} \label{eq:bc}
\begin{eqnarray}
\Phi(x^\prime,0) = \Phi(x^\prime,D/\cos\theta), 
\label{eq:bca}\\
\frac{\partial}{\partial z^\prime}\Phi(x^\prime,z^\prime)|_{z^\prime=0} = 
\frac{\partial}{\partial z^\prime}\Phi(x^\prime,z^\prime)|_{z^\prime=D/\cos\theta}, \label{eq:bcb}\\
\Phi(x^\prime,z^\prime)|_{x^\prime\rightarrow\pm\infty} = 0.
\label{eq:bcc}
\end{eqnarray}
\end{mathletters}

\section{Numerical procedures}

\subsection{Procedure I}

Taken into account the boundary conditions, Eq.~\ref{eq:eq2DFinal} can be represented as 
\begin{eqnarray}
\textbf{U}\bbox{\Phi}=\textbf{0}, 
\label{eq:UPhi}
\end{eqnarray}
where the column vector \mathversion{bold} $\Phi$ \mathversion{normal} $ = 
\{\Phi_k\}^\prime, \ k=1,2,...,(2n-2)\times2n$ represents the discrete solutions of Eq.~\ref{eq:eq2DFinal} (here $^\prime$ indicates transpose, $k$ and $n$ are integers).  
The sparse matrix \textbf{U} has the following structure, 
\blong
\bea
\textbf{U}=\left(
\matrix{
\textbf{U}_{3,2} &\textbf{U}_{4,2} &\textbf{U}_{5,2} &    &    &    &     &     &     \cr
\textbf{U}_{2,3} &\textbf{U}_{3,3} &\textbf{U}_{4,3} &\textbf{U}_{5,3} & & & & &      \cr
\textbf{U}_{1,4} &\textbf{U}_{2,4} &\textbf{U}_{3,4} &\textbf{U}_{4,4} &\textbf{U}_{5,4}   
&       &     &     &     \cr
    &\textbf{U}_{1,5} &\textbf{U}_{2,5} &\textbf{U}_{3,5} &\textbf{U}_{4,5}    &\textbf{U}_{5,5}    &     &     &     \cr
    &    &    &    &\ddots &       &     &     &     \cr
    &    &    &\textbf{U}_{1,2n-4} &\textbf{U}_{2,2n-4}    &\textbf{U}_{3,2n-4}    &\textbf{U}_{4,2n-4}  &\textbf{U}_{5,2n-4}  &     \cr
    &    &    &    &\textbf{U}_{1,2n-3}    &\textbf{U}_{2,2n-3}    &\textbf{U}_{3,2n-3}  &\textbf{U}_{4,2n-3}  &\textbf{U}_{5,2n-3}  \cr
    &    &    &    &       &\textbf{U}_{1,2n-2}    &\textbf{U}_{2,2n-2}  &\textbf{U}_{3,2n-2}  &\textbf{U}_{4,2n-2}  \cr      
    &    &    &    &       &       &\textbf{U}_{1,2n-1}  &\textbf{U}_{2,2n-1}  &\textbf{U}_{3,2n-1}  \cr
}
\right),
\label{eq:U}
\eea
\elong\noindent
where the block matrices {\bf U}$_{l,i}$ ($l=1,2,3,4,5$) can be expressed as follows 
\blong
\bea
\textbf{U}_{1,i}=
\left(
{\scriptsize
\matrix{
0 & & &    &       &       &  &  &0     \cr
c_{2,i,2} &c_{3,i,2} &c_{4,i,2} &c_{5,i,2} &       &       &     &c_{1,i,2}  &   \cr
c_{1,i,3} &c_{2,i,3} &c_{3,i,3} &c_{4,i,3} &c_{5,i,3}    &       &     &     &   \cr
    &c_{1,i,4} &c_{2,i,4} &c_{3,i,4} &c_{4,i,4}    &c_{5,i,4}    &     &     &   \cr
    &    &    &    &\ddots &       &     &     &     \cr
    &    &    &c_{1,i,2n-3}  &c_{2,i,2n-3}     &c_{3,i,2n-3}     &c_{4,i,2n-3}   
&c_{5,i,2n-3}   &     \cr
    &    &    &    &c_{1,i,2n-2}     &c_{2,i,2n-2}     &c_{3,i,2n-2}   &c_{4,i,2n-2}   &c_{5,i,2n-2}   \cr
    &c_{5,i,2n-1} &    &    &       &c_{1,i,2n-1}     &c_{2,i,2n-1}   &c_{3,i,2n-1}   &c_{4,i,2n-1}   \cr      
0    & & &    &       &       &  &  &0   \cr
}
}
\right),
\label{eq:U1}
\eea

\bea
\textbf{U}_{2,i}=
\left(
{\scriptsize
\matrix{
0 & & &    &       &       &  &  &0     \cr
c_{7,i,2} &c_{8,i,2} &c_{9,i,2} &c_{10,i,2} &       &       &     &c_{6,i,2}  &   \cr
c_{6,i,3} &c_{7,i,3} &c_{8,i,3} &c_{9,i,3} &c_{10,i,3}    &       &     &     &   \cr
    &c_{6,i,4} &c_{7,i,4} &c_{8,i,4} &c_{9,i,4}    &c_{10,i,4}    &     &     &   \cr
    &    &    &    &\ddots &       &     &     &     \cr
    &    &    &c_{6,i,2n-3}  &c_{7,i,2n-3}     &c_{8,i,2n-3}     &c_{9,i,2n-3}   
&c_{10,i,2n-3}   &     \cr
    &    &    &    &c_{6,i,2n-2}     &c_{7,i,2n-2}     &c_{8,i,2n-2}   &c_{9,i,2n-2}   &c_{10,i,2n-2}   \cr
    &c_{10,i,2n-1} &    &    &       &c_{6,i,2n-1}     &c_{7,i,2n-1}   &c_{8,i,2n-1}   &c_{9,i,2n-1}   \cr      
0    & & &    &       &       &  &  &0   \cr
}
}
\right),
\label{eq:U2}
\eea

\bea
\textbf{U}_{3,i}=
\left(
{\scriptsize
\matrix{
1      &       &       &       &       &       &       &       &-1      \cr
c_{12,i,2} &c_{13,i,2} &c_{14,i,2} &c_{15,i,2} &       &       &     &c_{11,i,2}  &   \cr
c_{11,i,3} &c_{12,i,3} &c_{13,i,3} &c_{14,i,3} &c_{15,i,3}    &       &     &     &   \cr
    &c_{11,i,4} &c_{12,i,4} &c_{13,i,4} &c_{14,i,4}    &c_{15,i,4}    &     &     &   \cr
    &    &    &    &\ddots &       &     &     &     \cr
    &    &    &c_{11,i,2n-3}  &c_{12,i,2n-3}     &c_{13,i,2n-3}     &c_{14,i,2n-3}   &c_{15,i,2n-3}   &     \cr
    &    &    &    &c_{11,i,2n-2}     &c_{12,i,2n-2}     &c_{13,i,2n-2}   &c_{14,i,2n-2}   &c_{15,i,2n-2}   \cr
    &c_{15,i,2n-1} &    &    &       &c_{11,i,2n-1}     &c_{12,i,2n-1}   &c_{13,i,2n-1}   &c_{14,i,2n-1}   \cr      
-6     &4     &-1     &       &       &       &-1     &4      &        \cr
}
}
\right),
\label{eq:U3}
\eea

\bea
\textbf{U}_{4,i}=
\left(
{\scriptsize
\matrix{
0 & & &    &       &       &  &  &0     \cr
c_{17,i,2} &c_{18,i,2} &c_{19,i,2} &c_{20,i,2} &       &       &     &c_{16,i,2}  &   \cr
c_{16,i,3} &c_{17,i,3} &c_{18,i,3} &c_{19,i,3} &c_{20,i,3}    &       &     &     &   \cr
    &c_{16,i,4} &c_{17,i,4} &c_{18,i,4} &c_{19,i,4}    &c_{20,i,4}    &     &     &   \cr
    &    &    &    &\ddots &       &     &     &     \cr
    &    &    &c_{16,i,2n-3}  &c_{17,i,2n-3}     &c_{18,i,2n-3}     &c_{19,i,2n-3}   &c_{20,i,2n-3}   &     \cr
    &    &    &    &c_{16,i,2n-2}     &c_{17,i,2n-2}     &c_{18,i,2n-2}   &c_{19,i,2n-2}   &c_{20,i,2n-2}   \cr
    &c_{20,i,2n-1} &    &    &       &c_{16,i,2n-1}     &c_{17,i,2n-1}   &c_{18,i,2n-1}   &c_{19,i,2n-1}   \cr      
0    & & &    &       &       &  &  &0   \cr
}
}
\right),
\label{eq:U4}
\eea

\bea
\textbf{U}_{5,i}=
\left(
{\scriptsize
\matrix{
0 & & &    &       &       &  &  &0     \cr
c_{22,i,2} &c_{23,i,2} &c_{24,i,2} &c_{25,i,2} &       &       &     &c_{21,i,2}  &   \cr
c_{21,i,3} &c_{22,i,3} &c_{23,i,3} &c_{24,i,3} &c_{25,i,3}    &       &     &     &   \cr
    &c_{21,i,4} &c_{22,i,4} &c_{23,i,4} &c_{24,i,4}    &c_{25,i,4}    &     &     &   \cr
    &    &    &    &\ddots &       &     &     &     \cr
    &    &    &c_{21,i,2n-3}  &c_{22,i,2n-3}     &c_{23,i,2n-3}     &c_{24,i,2n-3}   &c_{25,i,2n-3}   &     \cr
    &    &    &    &c_{21,i,2n-2}     &c_{22,i,2n-2}     &c_{23,i,2n-2}   &c_{24,i,2n-2}   &c_{25,i,2n-2}   \cr
    &c_{25,i,2n-1} &    &    &       &c_{21,i,2n-1}     &c_{22,i,2n-1}   &c_{23,i,2n-1}   &c_{24,i,2n-1}   \cr      
0    & & &    &       &       &  &  &0   \cr
}
}
\right).
\label{eq:U5}
\eea
\elong \noindent

The elements $c_{1,i,j}$, $c_{2,i,j}$, ..., $c_{25,i,j}$ are the coefficients of 
the discretized equations of Eq.~\ref{eq:eq2DFinal}. Considering Eq.~\ref{eq:U}, 
the ranges of the index $i$ for \textbf{U}$_{1,i}$, \textbf{U}$_{2,i}$, \textbf{U}$_{3,i}$, \textbf{U}$_{4,i}$ and \textbf{U}$_{5,i}$ are $4 \rightarrow 2n-1$, $3 \rightarrow 2n-1$, 
$2 \rightarrow 2n-1$, $2 \rightarrow 2n-2$ and $2 \rightarrow 2n-3$, respectively. 
The range of the index $j$ for the elements in \textbf{U}$_{l,i}$ is from 2 to $2n-1$. 
We can find the dimension of \textbf{U} as follows: each matrix of the form 
\textbf{U}$_{3,i}$ is of order 2n; the main diagonal consists of matrices of 
the form \textbf{U}$_{3,i}$ and there are $2n-2$ of them (the range of the index $i$ 
for \textbf{U}$_{3,i}$ is $2 \rightarrow 2n-1$), the dimension of \textbf{U} is thus 
$(2n-2)2n\times(2n-2)2n$.

The boundary conditions of Eq.~\ref{eq:bc} have been incorporated into $\textbf{U}$: 
the periodic condition of Eq.~\ref{eq:bca} is explicitly expressed in 
all the first rows of \textbf{U}$_{3,i}$ (see Eq.~\ref{eq:U3});  
the derivative condition of Eq.~\ref{eq:bcb} is treated by the four-point difference 
techniques in all the last rows of \textbf{U}$_{3,i}$; 
the zero condition of Eq.~\ref{eq:bcc} is implemented at $i=1,2n$, which is 
the reason that the range of the index $i$ for \textbf{U}$_{3,i}$ is from 2 to $2n-1$. 
The zero solutions on and outside the boundaries at $i=1,2n$ also lead to 
the structures of the first and last two rows in  \textbf{U} (see Eq.~\ref{eq:U}). 

Except for the first and last rows, all the other rows in \textbf{U}$_{l,i}$ have  
the same structure. In the second and second last rows of \textbf{U}$_{l,i}$, 
the periodic property of the solutions is considered. As exemplified in 
the second last rows of \textbf{U}$_{l,2n-1}$ ($l=1,2,3,4,5$), the five 
coefficients $c_{5,2n-1,2n-1}$, $c_{10,2n-1,2n-1}$, $c_{15,2n-1,2n-1}$, $c_{20,2n-1,2n-1}$ 
and $c_{25,2n-1,2n-1}$ are there (cf. Eqs.~\ref{eq:U1}-\ref{eq:U5}) 
 because the discrete solutions corresponding to these coefficients 
are equivalent to the counterparts at $j=2$. 

For non-trivial solutions, the determinant of $\textbf{U}$ should be zero, 
\begin{eqnarray}
\det|\textbf{U}|=0. 
\label{eq:detU}
\end{eqnarray}
By eliminating the constant elements in the first and last rows (columns) in the matrices $\textbf{U}_{3,i}$, namely, eliminating the first and last rows (columns) in $\textbf{U}_{l,i}$, Eq.~\ref{eq:detU} can be transformed into 
\begin{eqnarray}
\det|\textbf{U}^\prime|=0, 
\label{eq:detUp}
\end{eqnarray}
where the block matrices $\textbf{U}_{l,i}^\prime$ ($l=1, 2, 3, 4, 5$ and $i$ is 
the counterpart subscript in $\textbf{U}_{l,i}$) in $\textbf{U}^\prime$ have 
the following general structure, 
\blong
\bea
\textbf{U}_{l,i}^\prime= 
\left(
{\scriptsize
\matrix{
\epsilon_{2}+\frac{2}{3}\nu_{2}   &\sigma_{2}-\frac{1}{6}\nu_{2}   &\tau_{2}   &     &     
&    &    &-\frac{1}{6}\nu_{2}    &\mu_{2}+\frac{2}{3}\nu_{2}                       \cr
\nu_{3} +\frac{2}{3}\mu_{3}       &\epsilon_{3}-\frac{1}{6}\mu_{3}    &\sigma_{3} 
&\tau_{3} &     &    &    &-\frac{1}{6}\mu_{3}     &\frac{2}{3}\mu_{3}              \cr
\mu_{4} &\nu_{4} &\epsilon_{4} &\sigma_{4}   &\tau_{4}   &      &      &     &      \cr
    &\mu_{5} &\nu_{5}      &\epsilon_{5} &\sigma_{5} &\tau_{5}     &       &    &   \cr
    &    &         &         &\ddots &         &         &       &     	            \cr
    &    &         &\mu_{2n-4}      &\nu_{2n-4}    &\epsilon_{2n-4} &\sigma_{2n-4}   &\tau_{2n-4}   &                           \cr
    &    &         &         &\mu_{2n-3}    &\nu_{2n-3}      &\epsilon_{2n-3} 
&\sigma_{2n-3} &\tau_{2n-3}                \cr
\frac{2}{3}\tau_{2n-2}         &-\frac{1}{6}\tau_{2n-2}    &      &     &     
&\mu_{2n-2} &\nu_{2n-2}  &\epsilon_{2n-2}-\frac{1}{6}\tau_{2n-2} 
&\sigma_{2n-2}+\frac{2}{3}\tau_{2n-2}      \cr      
\tau_{2n-1}+\frac{2}{3}\sigma_{2n-1}    &-\frac{1}{6}\sigma_{2n-1}    &      &     &     
&    &\mu_{2n-1}  &\nu_{2n-1}-\frac{1}{6}\sigma_{2n-1}    
&\epsilon_{2n-1}+\frac{2}{3}\sigma_{2n-1}  \cr
}
}
\right).
\label{eq:Up}
\eea
\elong\noindent
As a matter of convenience, the notations $l, i$ are omitted in the elements above and 
only the index $j$ ($j=2,3,...,2n-1$) is given. The quantities of $\mu(l)_{i,j}$, 
$\nu(l)_{i,j}$, $\epsilon(l)_{i,j}$, $\sigma(l)_{i,j}$ and $\tau(l)_{i,j}$ indicate 
the corresponding elements from \textbf{U}. For example, $\epsilon_{j}$ in $\textbf{U}_{3,i}^\prime$ is $\epsilon(3)_{i,j} \leftrightarrow c_{13,i,j}$ in $\textbf{U}_{3,i}$; $\mu_4$ in $\textbf{U}_{1,i}^\prime$ is $\mu(1)_{i,4} 
\leftrightarrow c_{1,i,4}$ in $\textbf{U}_{1,i}$; $\mu_4$ in $\textbf{U}_{2,i}^\prime$ 
is $\mu(2)_{i,4} \leftrightarrow c_{6,i,4}$ in $\textbf{U}_{2,i}$; $\mu_4$ in $\textbf{U}_{3,i}^\prime$ is $\mu(3)_{i,4} \leftrightarrow c_{11,i,4}$ in 
$\textbf{U}_{3,i}$. Each $\textbf{U}_{l,i}^\prime$ is of order $2n-2$. Thus, 
the dimension of $\textbf{U}^\prime$ is $(2n-2)(2n-2)\times(2n-2)(2n-2)$, 
which is less than that of \textbf{U}. 

It can be verified that 
the parameters of the magnetic field $B$, contained in the $c_{13,i,j}$ coefficients, 
only appear in the main diagonal position of $\textbf{U}$ (thus $\textbf{U}^\prime$). 
Since $c_{13,i,j}$ can be categorized into two different parts: with and 
without $B$, Eq.~\ref{eq:detUp} can be further written as 
\begin{eqnarray}
\det|\textbf{P}-B^{2}\textbf{I}|=0, 
\label{eq:detP}
\end{eqnarray}
where $\textbf{I}$ is a unitary matrix. \textbf{P} has the same structure and 
dimension as $\textbf{U}^\prime$ and the block matrices $\textbf{P}_{l,i}$ in \textbf{P} 
have the following general structure, 
\blong
\bea
\textbf{P}_{l,i}= 
\left(
{\scriptsize
\matrix{
\epsilon^\prime_{2}+\frac{2}{3}\nu_{2}   &\sigma_{2}-\frac{1}{6}\nu_{2}   &\tau_{2}   
&     &     &    &    &-\frac{1}{6}\nu_{2}          &\mu_{2}+\frac{2}{3}\nu_{2}         \cr
\nu_{3} +\frac{2}{3}\mu_{3}       &\epsilon^\prime_{3}-\frac{1}{6}\mu_{3}    &\sigma_{3} &\tau_{3} &     &    &    &-\frac{1}{6}\mu_{3}          &\frac{2}{3}\mu_{3}             \cr
\mu_{4} &\nu_{4} &\epsilon^\prime_{4} &\sigma_{4}   &\tau_{4}   &     &    &     &      \cr
    &\mu_{5} &\nu_{5}      &\epsilon^\prime_{5} &\sigma_{5} &\tau_{5}     &   &  &      \cr
    &    &         &         &\ddots &         &         &       &      	          \cr
    &    &         &\mu_{2n-4}      &\nu_{2n-4}    &\epsilon^\prime_{2n-4} &\sigma_{2n-4}   &\tau_{2n-4}   &                                  \cr
    &    &         &         &\mu_{2n-3}    &\nu_{2n-3}      &\epsilon^\prime_{2n-3} &\sigma_{2n-3} &\tau_{2n-3}                       \cr
\frac{2}{3}\tau_{2n-2}      &-\frac{1}{6}\tau_{2n-2}    &      &     &     &\mu_{2n-2} &\nu_{2n-2}  &\epsilon^\prime_{2n-2}-\frac{1}{6}\tau_{2n-2} 
&\sigma_{2n-2}+\frac{2}{3}\tau_{2n-2}             \cr      
\tau_{2n-1}+\frac{2}{3}\sigma_{2n-1}    &-\frac{1}{6}\sigma_{2n-1}   &      &     &     &    &\mu_{2n-1}  &\nu_{2n-1}-\frac{1}{6}\sigma_{2n-1}    
&\epsilon^\prime_{2n-1}+\frac{2}{3}\sigma_{2n-1}  \cr
}
}
\right), 
\label{eq:Pl}
\end{eqnarray} 
where $\epsilon_{j}^\prime$ is short for $\epsilon^\prime(l)_{i,j}$ and 
one has 
\begin{eqnarray} 
\epsilon^\prime(l)_{i,j}=\left\{ \begin{array}{ll}
\epsilon(l)_{i,j} & l=1,2,4,5 \\
\epsilon(l)_{i,j}-B^2 & l=3. \end{array} \right.
\label{eq:relation}
\eea
\elong\noindent
Thus, $\textbf{P}_{l,i}|_{l=1,2,4,5} = \textbf{U}_{l,i}^\prime|_{l=1,2,4,5}$ but 
$\textbf{P}_{3,i} \not= \textbf{U}_{3,i}^\prime$.

From Eq.~\ref{eq:detP}, it is clear that the largest solution for $B$, namely $B_{c2}$, 
can be obtained from the maximum eigenvalue of the following eigen equation,
\begin{eqnarray}
\textbf{P}\bbox{\chi}=B^{2}\bbox{\chi},
\label{eq:PChi}
\end{eqnarray}
where \mathversion{bold}{$\chi$} \mathversion{normal} is the eigen function of 
$\textbf{P}$. Having obtained $B_{c2}$, the corresponding order parameter can be 
obtained by substituting $B_{c2}$ into Eq.~\ref{eq:UPhi}.

\subsection{Procedure II}

In the above procedure, the magnetic field square $B^2$ has been treated as 
an eigenvalue (Eq.~\ref{eq:PChi}). In fact, one can directly discretize 
Eq.~\ref{eq:eq2DFinal} into a matrix eigen equation, from which the upper critical field 
can be directly deduced, 
\begin{eqnarray}
\textbf{Q}\bbox{\Phi}=B^{2}\bbox{\Phi}, 
\label{eq:QPhi}
\end{eqnarray}
where the matrix \textbf{Q} has the same structure as \textbf{U} and the block matrices 
in \textbf{Q} have the following general structure, 
\blong
\bea
\textbf{Q}_{l,i}=
\left(
{\small
\matrix{
\epsilon^\prime_{1}  &\sigma_{1}  &\tau_{1}    &       &       &  &\mu_{1}  &\nu_{1}  &  \cr
\nu_{2} &\epsilon^\prime_{2} &\sigma_{2} &\tau_{2} &       &       &     &\mu_{2}  &     \cr
\mu_{3} &\nu_{3} &\epsilon^\prime_{3} &\sigma_{3} &\tau_{3}    &       &     &     &     \cr
    &\mu_{4} &\nu_{4} &\epsilon^\prime_{4} &\sigma_{4}    &\tau_{4}    &     &     &     \cr
    &    &    &    &\ddots &       &     &     &     \cr
    &    &    &\mu_{2n-3} &\nu_{2n-3} &\epsilon^\prime_{2n-3} &\sigma_{2n-3} &\tau_{2n-3} &  \cr
    &    &    &    &\mu_{2n-2} &\nu_{2n-2} &\epsilon^\prime_{2n-2} &\sigma_{2n-2} &\tau_{2n-2}  \cr
    &\tau_{2n-1} &    &    &       &\mu_{2n-1}     &\nu_{2n-1}   &\epsilon^\prime_{2n-1}   &\sigma_{2n-1}   \cr      
    &\sigma_{2n}     &\tau_{2n}     &       &       &       &\mu_{2n}     &\nu_{2n}      &\epsilon^\prime_{2n}        \cr
}
}
\right).
\label{eq:Q}
\eea
\elong\noindent
Here, the quantities of $\mu_{j}$, $\nu_{j}$, $\epsilon^\prime_{j}$, $\sigma_{j}$ and $\tau_{j}$ have the same meaning as those in $\textbf{P}_{l,i}$. Unlike that in $\textbf{U}_{l,i}$, $\textbf{U}_{l,i}^\prime$ and $\textbf{P}_{l,i}$, the range of 
the index $j$ in $\textbf{Q}_{l,i}$ is now from $1$ to $2n$. The dimension of 
\textbf{Q} is the same as that of \textbf{U}, i.e., $(2n-2)2n\times(2n-2)2n$. 
Note that the periodic property of the solutions has been considered implicitly in 
the first and last rows of $\textbf{Q}_{l,i}$, unlike the explicit condition in 
$\textbf{U}_{3,i}$ (see Eq.~\ref{eq:U3}). Such a periodic property also appears in 
the second and last second rows in $\textbf{Q}_{l,i}$ as well as in $\textbf{U}_{l,i}$. 

The present procedure allows a direct determination of the upper critical field 
from the largest eigenvalue of the magnetic field square of the linear GL equation, 
compared to the conventional method that $B_{c2}$ is obtained via the lowest eigenvalue of 
the Landau level of the linear GL equation\cite{Koyama92,Abrikosov57,Tinkham96,Cheng99,Joynt90,Berlinsky95,Franz96,Chang98}. 

\section{Results and Discussion}

In the above procedures, we have treated the magnetic field square as 
eigenvalue problems (Eq.~\ref{eq:PChi} and Eq.~\ref{eq:QPhi}) and hence, 
the upper critical field in the $c$-$a$ plane can be directly obtained from 
the corresponding eigen equations. It should be noted that these procedures are 
related to the partial differential Eq.~\ref{eq:eq2DFinal}, namely a 
2D differential equation. Note that we can have similar procedures to 
determine the upper critical field parallel to the layers\cite{Wang01b} but  
these procedures (corresponding to Eqs.~5 and 6 in Ref.~\onlinecite{Wang01b}, respectively)   
are now related to an ordinary differential equation, i.e., a 1D differential equation 
(see Eq.~1 in Ref.~\onlinecite{Wang01b}). The values of $B_{c2}$ of Bi2212, calculated from 
the two 2D procedures under the conditions of $\theta=89.9^\circ$ and $T=0$ K, are 
presented in Table~\ref{tab:Bi2212Bc22D}. The corresponding 1D calculations 
at $\theta=90^\circ$ and $T=0$ K are also listed. It can be seen that 
the results obtained from the 2D procedures ($\theta = 89.9^\circ$) are in 
good agreement with each other and reasonably approximate the corresponding 
1D calculations ($\theta = 90^\circ$). 

It is found that the order parameter obtained at 
$B_{c2}$  from procedure I (Eq.~\ref{eq:PChi}) is also 
consistent with that from procedure II (Eq.~\ref{eq:QPhi}) and we shall utilize procedure II  for the following calculations. 
In Fig.~1, we show the surface and contour plots of the order parameter 
for Bi2212 with the conditions $\theta=89.9^\circ$ and $T= 0$ K. Note that 
the $x^\prime$-axis nearly overlaps the $z$- ($c$-) axis 
(namely, $x^\prime \approx z$) at $\theta=89.9^\circ$. Hence, as depicted in Fig.~1, the order parameter of Bi2212 is localized in 
a thin slab ($z \in [-2,2]$ a.u.) along the $c$- axis. The behavior found here is 
consistent with the fact that for Bi2212 the $c$-axis coherence length at 
zero temperature is very short\cite{Han98} and less than the distance between 
the two effective superconducting layers ($\sim 12.34$ \AA, see Ref.~\onlinecite{Wang01b}). 
However, for another intrinsically layered superconductor YBCO with 
relatively small anisotropy\cite{Poole00}, we found that the localized domain at 
zero temperature is about $z \in [-10,10]$ a.u., which is broader than the distance 
between the two effective superconducting layers ($\sim 8.41$ \AA, see Ref.~\onlinecite{Wang01b}). 
As such, one may say that at zero temperature, Bi2212 exhibits a 2D feature while 
YBCO demonstrates a quasi-3D behavior. 

Fig.~2 shows the order parameter 
and its contour at $B_{c2}$ for Bi2212 at a higher temperature 0.9 $T_c$, with 
the angle $\theta=89.9^\circ$ unchanged. It is obvious that the order parameter 
becomes broader than the case at zero temperature (Fig.~1), 
as expected. The localized domain across the layer is now about $z \in 
[-6,6]$ a.u. and it is still smaller than the distance between the adjacent superconducting layers. 
This behavior is qualitatively consistent with the 2D feature of Bi2212 at 0.9 $T_c$ 
calculated by the one-dimensional procedure. 
However, for YBCO, we find that the localized domain at 0.9 $T_c$ is about $z \in 
[-15,15]$ a.u., larger again than the distance between the nearest superconducting layers. Consequently, one may conclude that Bi2212 is a 2D superconductor in a large temperature 
while YBCO is intrinsically a quasi-3D superconductor. Note that the order parameter 
presented in Figs.~1 and 2 are found to be of 
a Gaussian type, which is consistent with the fact that the profile of the order parameter 
associated with the smallest eigenvalue (Landau level) of the usual linear GL equation 
at $B_{c2}$. 

Finally, it is worth mentioning that Eq.~\ref{eq:PChi} of procedure I requires 
less memory than Eq.~\ref{eq:QPhi} of procedure II in obtaining $B_{c2}$ since 
the dimension of \textbf{P} is less than that of $\textbf{Q}$. The number of 
the stored elements in the sparse matrix \textbf{P} is less than that in 
\textbf{Q} by $(2n-2)2n \times (2n-2)2n-(2n-2)(2n-2) \times (2n-2)(2n-2) = 
16(2n-1)(n-1)^2$. When $n$ is large, the reduced amount ($\approx 32n^3$) is 
significant. However, in order to obtain both $B_{c2}$ and the associated order 
parameter, procedure I requires more CPU time than procedure II since in 
using procedure I, two equations (Eqs.~\ref{eq:UPhi} and \ref{eq:PChi}) have to 
be solved while only one equation (Eq.~\ref{eq:QPhi}) is involved in procedure II.

\section{Conclusion}

In this paper, we applied our continuous Ginzburg-Landau model to 
layered superconductors in the presence of a magnetic field parallel to 
the $c$-$a$ plane. By treating the magnetic field square as eigenvalues, two procedures 
were proposed to determine the upper critical field. Near the direction parallel to 
the layers, the critical fields of Bi2212 calculated from the two procedures are 
consistent with each other and are good approximations to the corresponding parallel 
upper critical fields determined by similar procedures. The behaviors of 
the order parameter obtained at $B_{c2}$ are reasonable. Our procedures can be 
a useful starting point for investigating the properties associated with 
the upper critical field and the corresponding order parameter.

\begin{table}
\caption{Values of $B_{c2}$ (Tesla) at 89.9$^\circ$ at 0 K for Bi2212, 
determined by the two 2D procedures (Eq.~\ref{eq:PChi} and Eq.~\ref{eq:QPhi}). 
The results from the 2D procedures are consistent with each other and 
reasonably approximate the parallel upper critical fields determined by 
the corresponding 1D procedures.}
\begin{tabular}{llc}
\ procedure    		        & n   &  $B_{c2}$     \\
\hline
				        & 30  &  3342.3065    \\
\ I (2D, Eq.~\ref{eq:PChi})       & 40  &  3337.3166   \\ 
					  & 50  &  3336.7392    \\
\hline
\ I (1D)           & 800 &  3338.6584    \\
\hline\hline
					  & 30  &  3338.7414    \\
\ II (2D, Eq.~\ref{eq:QPhi})      & 40  &  3336.9549   \\ 
					  & 50  &  3336.8420   \\
\hline
\ II (1D)          & 800 &  3338.6801    \\
\end{tabular} \label{tab:Bi2212Bc22D}
\end{table}

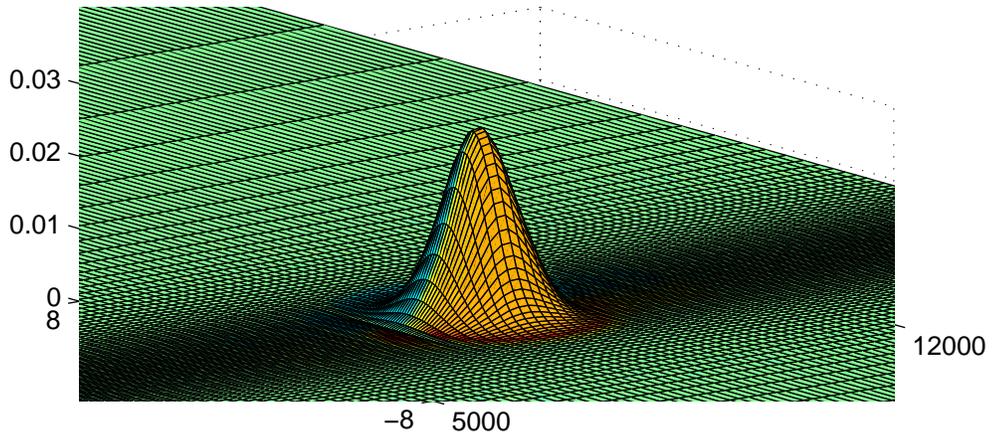
\begin{figure}[hbtp]
\caption[]{
Surface and contour plots of the order parameter of Bi2212 at zero temperature. 
The plots are depicted against the $z^\prime-x^\prime$ plane ($x^\prime$ is 
from negative to positive and $z^\prime$ is from 5000 to 12000). The order parameter 
is localized in a narrow domain (about 6. a.u.) along the $c$-axis unit cell.
}
\end{figure}

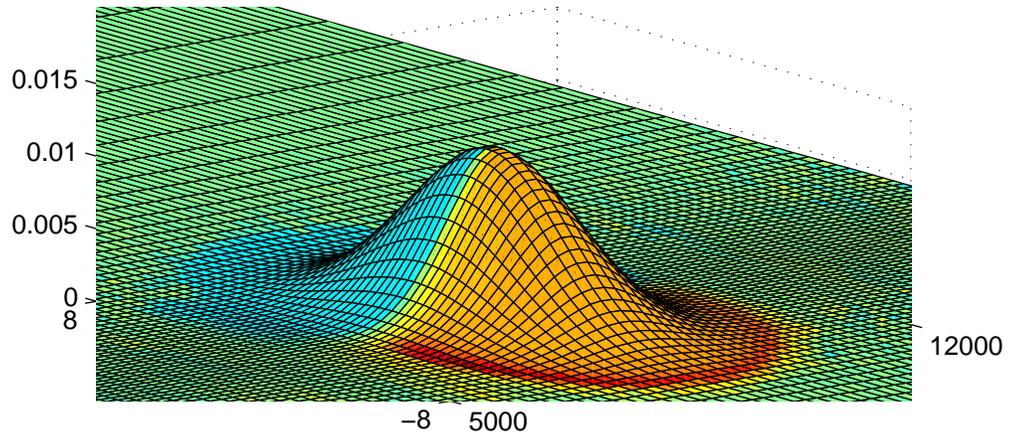
\begin{figure}[hbtp]
\caption[]{
Surface and contour plots of the order parameter of Bi2212 at 0.9 $T_c$. 
The order parameter spreads out and becomes broader than that at zero temperature 
(see Fig.~1).
}
\end{figure}

\end{multicols}
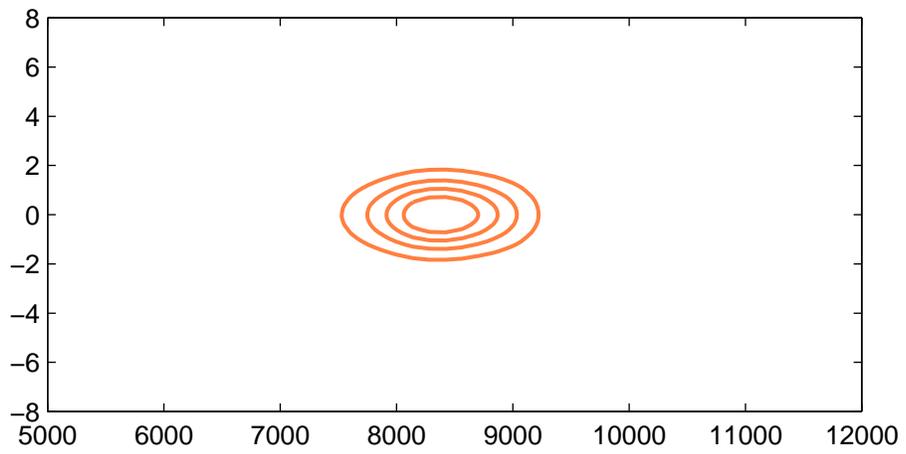
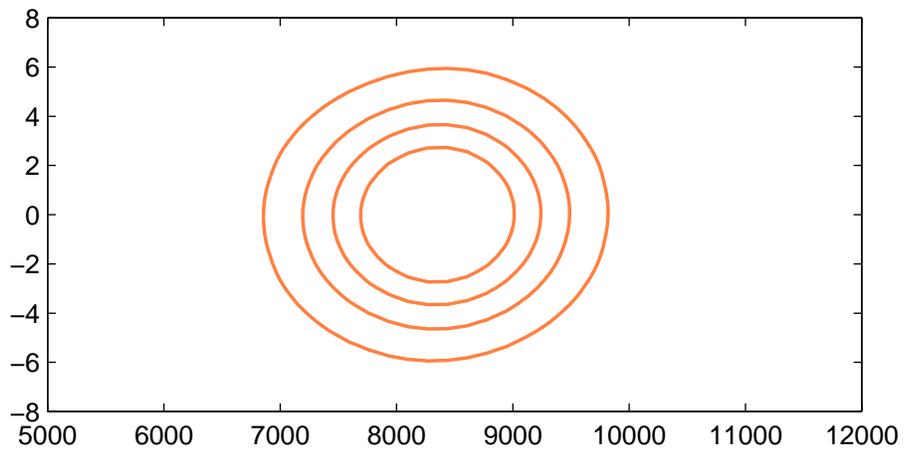
\end{document}